\documentclass[aps,prl,twocolumn,showpacs]{revtex4}
\usepackage{amssymb}
\usepackage{amsmath}
\usepackage{epsfig}
\usepackage{color}
\newcommand{\sech}{\rm sech}

\begin{document}

\title{All-phononic Digital Transistor on the Basis of Gap-Soliton Dynamics in Anharmonic Oscillator Ladder}
\author {Merab Malishava, Ramaz Khomeriki}
\affiliation {Physics Department, Javakhishvili Tbilisi State
University, 3 Chavchavadze, 0179 Tbilisi, Georgia}
\begin{abstract}
A conceptual mechanism of amplification of phonons by phonons on the
basis of nonlinear band-gap transmission (supratransmission)
phenomenon is presented. As an example a system of weakly coupled
chains of anharmonic oscillators is considered. One (source) chain
is driven harmonically by boundary with a frequency located in the
upper band close to the band edge of the ladder system.
Amplification happens when a second (gate) chain is driven by a
small signal in the counter phase and with the same frequency as
first chain. If the total driving of both chains overcomes the
band-gap transmission threshold the large amplitude band-gap soliton
emerges and amplification scenario is realized. The mechanism is
interpreted as nonlinear superposition of evanescent and propagating
nonlinear modes manifesting in a single or double soliton generation
working in band-gap or band-pass regimes, respectively. The results
could be straightforwardly generalized for all-optical or
all-magnonic contexts and has all the promises for logic gate
operations.
\end{abstract}

 \maketitle

Since the celebrated Fermi-Pasta-Ulam (FPU) first numerical
experiment \cite{fpu} in 1954, anharmonic oscillator chains became a
powerful tool in dealing with both fundamental aspects of
statistical physics \cite{izrailev, ruffo1} and nonlinear wave
phenomena \cite{scott} and, at the same time, serve as the simplest
prototypes for extremely complex condensed matter systems
\cite{flach1,flach2} and even biophysical processes
\cite{takeno,thierry2}. In particular, studies on the FPU chains
together with its further developments, namely nonintegrable
(Klein-Gordon \cite{klein} and Frenkel-Kontorova \cite{braun}) and
integrable Toda \cite{toda} chains, had an impact on the discovery
of solitons \cite{zabusky,thierry1} , helped much in understanding
of interplay between integrability and chaos \cite{chaos}, have been
widely applied for understanding of anomalous thermal conduction and
rectification properties in realistic physical systems
\cite{lepri1,peyrard,casati,casati1}, have applied to describe
transport properties in electric transmission lines \cite{trans} and
even in quantum systems, such as Josephson junction parallel arrays
and lattices \cite{zolot,ramaz3}, and untill now are widely used to
resolve thermal equipartition issues \cite{ponno}.

Surprisingly, the ladder extension of anharmonic one-dimensional
systems is rarely studied (but see Refs. \cite{book,kos}), although
there exists a wide range of applications for realistic systems,
e.g. optical directional couplers \cite{phot1,phot2}, weakly coupled
classical \cite{sievers} or quantum \cite{quant} spin chains,
coupled two or multicomponent systems \cite{ber,multi}, etc. In the
present letter we aim to consider two weakly coupled FPU chains in
order to realize digital all-phononic amplification of acoustic
signals. The considered concept of amplification could be
straightforwardly extended in case of similar all-optical
\cite{l1,l2} and all-magnonic \cite{magn1,magn2} devices.
\begin{figure}[b]
\includegraphics[scale=.33]{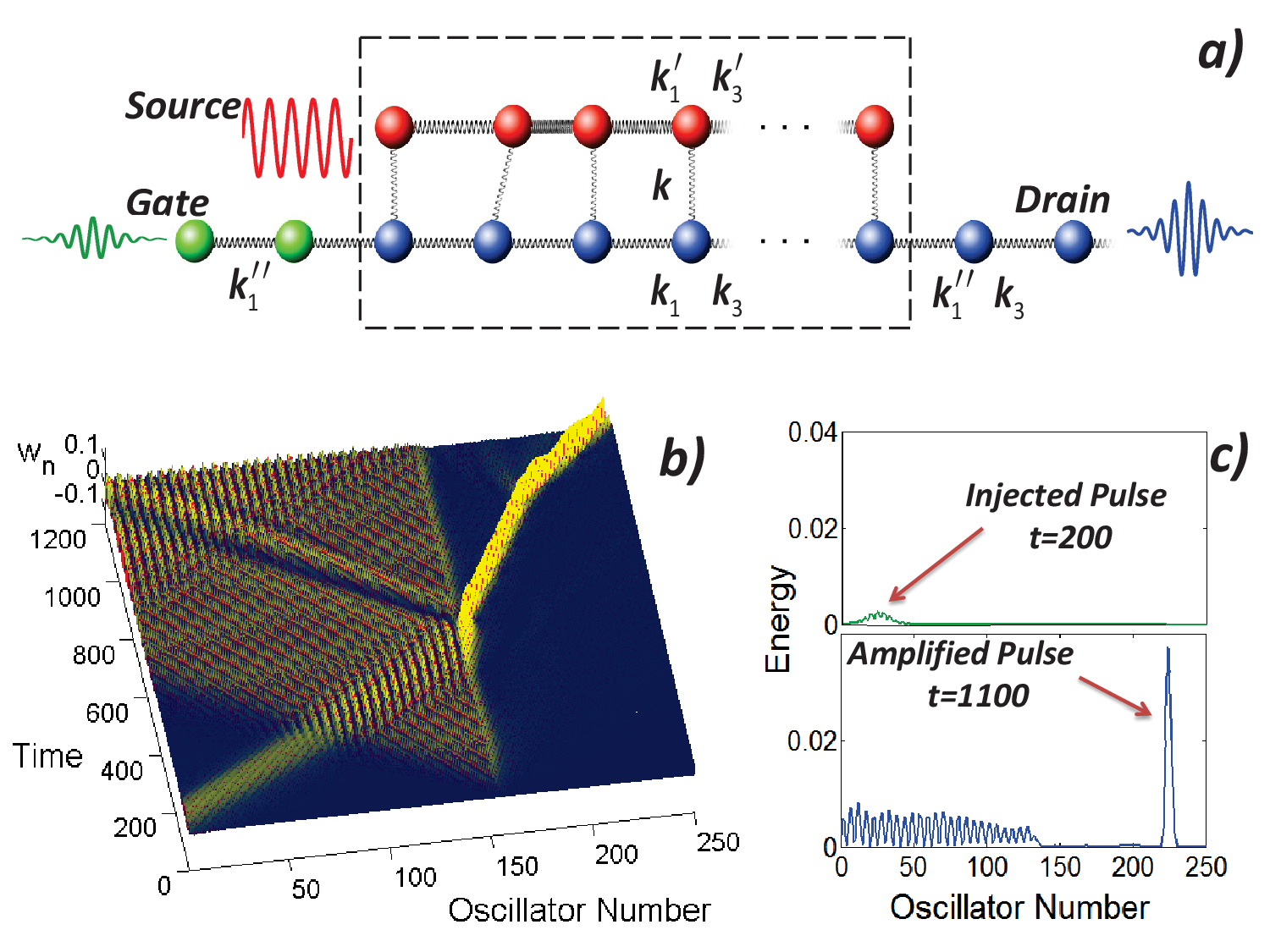}
\caption{a) The conceptual scheme of phonon transistor with
indications of source and gate signal forms and supply places.
Amplified signal is monitored at the Drain port. b) Space-time
dynamics of gap-soliton creation and propagation in the lower chain,
where in the range $1<n<150$ one has a linear chain (green balls in
the upper graph) and for $150<n<250$ soliton propagation occurs in a
nonlinear chain (blue balls). The upper chain is coupled with the
lower one in the range $150<n<200$. c) Energies of the input Gate
signal (green curve) and the amplified gap-soliton output Drain
signal (blue curve) monitored in the same lower chain}
\label{fig_01}
\end{figure}

Phonon laser \cite{laser,laser1} developments renew the interest in
various applications of monochromatic acoustic waves. Several ideas
have been proposed for phonon diodes
\cite{diode,diode1,diode2,diode3,mag,cry1,cry2} and all-phononic
transistors \cite{all,acoust} working on magneto-acoustic, nonlinear
wave-mixing or mode-mode interaction effects.

In this letter we implement a nonlinear band-gap transmission
mechanism \cite{leon,ramaz1,ramaz2} producing gap-solitons in order
to achieve digital amplification of weak acoustic signals. All three
ports of the proposed device work on a single operational frequency
and the schematics is presented in Fig. 1a, where the amplifying
part of the device is indicated by a dashed frame. Green and blue
color chains outside the frame are used for supplying the signal at
the gate and monitoring the output pulse at the Drain ports,
respectively. The signal is injected from the left (gate) linear
oscillator chain (green balls). The gate signal has a carrier
frequency within the band gap of the system of the nonlinear
oscillator ladder (red and blue balls inside the frame) and without
source driving cannot propagate further. At the source input we
apply large amplitude harmonic driving with the same frequency as a
gate signal. The amplitude of the source is just below the band gap
transmission threshold for antisymmetric mode of the ladder and
together with the gate signal the overall amplitude is enough to
exceed the threshold and a single large amplitude soliton passes the
ladder system and appears at the drain. While without the gate
signal the soliton is not produced. Thus it is clear that the
soliton amplitude is mostly defined by source driving and the
digital amplification scenario takes place. Main results of the
numerical simulations are presented in Fig. 1b and c). Particularly,
graph (b) describes space-time evolution of the displacements $w_n$
of the lower chain, while in graph (c) the energies of signals at
the Gate and Drain ports are displayed. For clarity we present a
movie file in Supplemental Material (SM) in order to show the signal
propagation and distribution among upper and lower chains.

For the analytical consideration we examine the ladder part of the
system (blue and red balls within the frame in Fig. 1a) modeling the
system as two weakly coupled FPU chains as follows:
\begin{eqnarray}
m\ddot u_n &=&k_1^\prime(u_{n+1}+u_{n-1}-2u_n)+k_3^\prime(u_{n+1}-u_n)^3 \nonumber \\
&&
+k_3^\prime(u_{n-1}-u_n)^3+k(w_n-u_n) \nonumber \\
m\ddot w_n &=&k_1(w_{n+1}+w_{n-1}-2w_n)+k_3(w_{n+1}-w_n)^3 \nonumber
\\ && +k_3(w_{n-1}-w_n)^3+k(u_n-w_n) \label{1}
\end{eqnarray}
where $u_n$ and $w_n$ stand for displacements of the $n$-th
oscillators (with mass $m$) of the upper and lower chains
respectively; $k$, $k_1^\prime$, $k_1$, $k_3$ and $k_3^\prime$ are
linear and nonlinear coefficients of stiffness of the springs.
Without restricting generality we rescale displacement amplitudes
and time such that the parameters of the upper chain take the unit
values $m=k_1^\prime=k_3^\prime=1$. All numerical simulations will
be done using this scaling and fixing the parameters of the lower
chain and interchain coupling as follows: $k_1=1.1$, $k_3=3.5$,
$k=0.2$. We apply dirichlet boundary condition at the left end of
both chains oscillating the balls $n=0$ of upper and lower chains
with the source and gate amplitudes, respectively.

In the linearized version of \eqref{1} we can readily define in the
$n$-th site of the ladder a two component vector $\left(u_n,
w_n\right)$ and seek for a solution in a form of harmonic waves
\begin{eqnarray}
\left(u_n, w_n\right)=\left(R,~~1\right)e^{i(pn-\Omega t)} +c.c.,
\label{2}
\end{eqnarray}
which gives us two branches of antisymmetric mode (neighboring the
$n$-th sites in different chains of the ladder oscillate in counter
phase) and symmetric mode (the $n$-th sites oscillate in the same
phase). Modes with corresponding dispersion relations $\Omega_1(p)$
and $\Omega_2(p)$ with maximum values at $p=\pi$ are displayed in
Fig. 2. Those modes are characterized by respective components $R_1$
and $R_2$ given by the following formulas
\begin{equation}
R_j=k/[k-(\Omega_j)^2+2(1-\cos p_j)]
\label{222}
\end{equation}
where $j=1,2$. In the case of identical chains in the ladder
$R_j=\pm1$, i.e. the oscillation amplitudes of interchain neighbor
oscillators are the same, while in our case of asymmetric ladder the
oscillation amplitudes are larger in the lower chain (for $j=1$) or
in the upper chain ($j=2$). If one drives the ladder with a
monochromatic frequency $\Omega$ the excitation wavenumbers $p_j$ of
the respective modes are calculated via the relations:
\begin{eqnarray}
\Omega=\Omega_1(p); \qquad \Omega=\Omega_2(p), \label{3}
\end{eqnarray}
\begin{figure}[b]
\includegraphics[scale=.35]{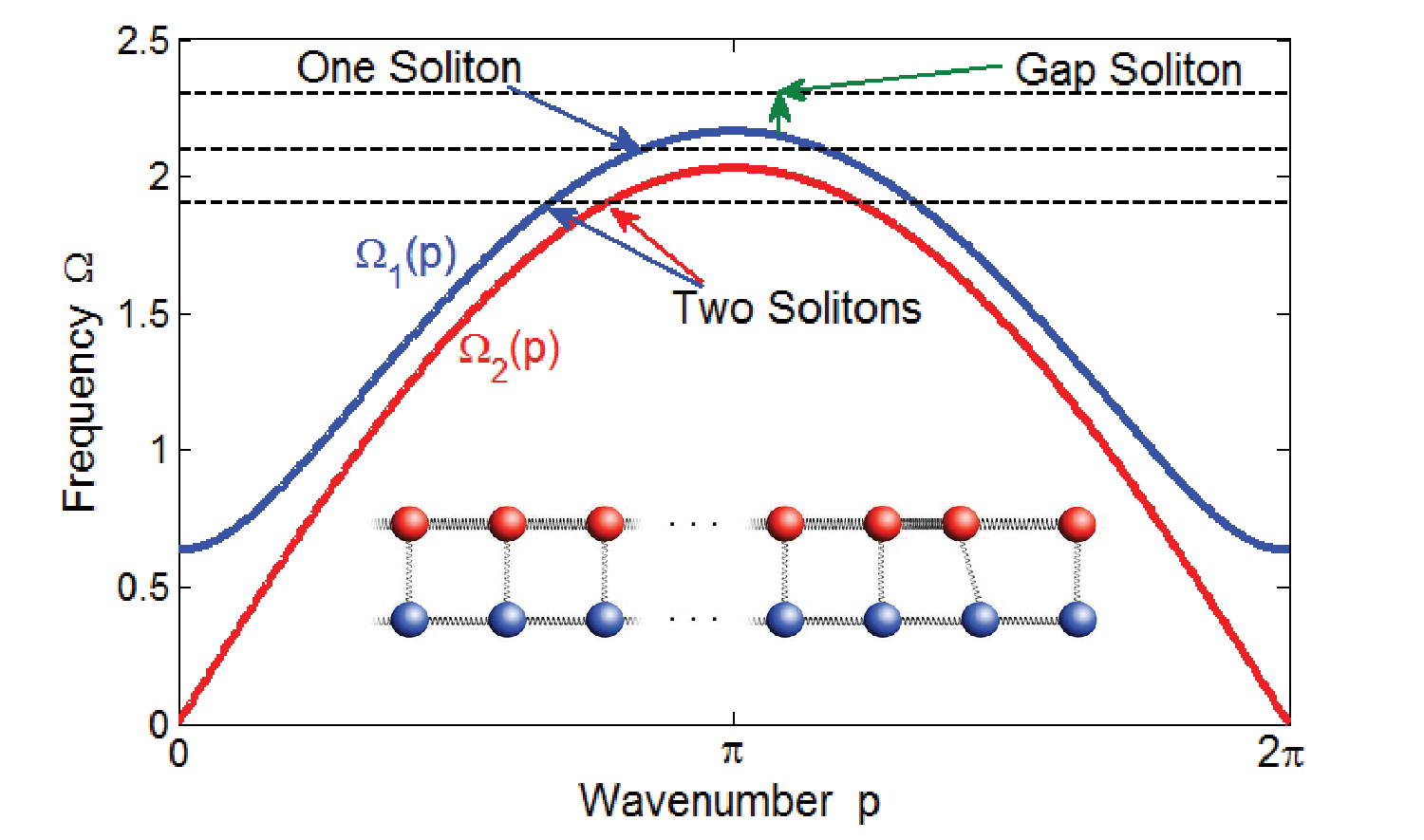}
\caption{Red and blue curves represent linear dispersion relations
of the antisymmetric $\Omega_1(p)$ and symmetric $\Omega_2(p)$
branches, respectively. Dashed horizontal lines represent three
different cases($\Omega=1.90$, $\Omega=2.10$ and $\Omega=2.18$)
which lead to the different kind of nonlinear wave transport in the
system. Inset shows the schematics of the system of the FPU ladder.}
\label{fig_2}
\end{figure}

Depending on the driving frequency (see Fig. 2) the following three
cases could be realized: 1) for excitation frequencies
$\Omega<\Omega_2(\pi)<\Omega_1(\pi)$ a nonlinear wave enters the
system, then separates into two soliton waves; 2) for excitation
frequencies $\Omega_2(\pi)<\Omega<\Omega_1(\pi)$ a nonlinear wave
generates a single soliton associated to the antisymmetric mode
$j=1$; And finally, 3) for the $\Omega>\Omega_1(\pi)>\Omega_2(\pi)$
a nonlinear wave cannot enter the system unless the driving
amplitude exceeds the band gap transmission threshold.

We start our analysis from considering driving frequencies, which
are within the band of both modes (the lowest dashed horizontal line
in Fig. 2). Then the wavenumbers $p_j$ of both modes are real and
could be found solving Eqs. \eqref{3}. Then a weakly nonlinear
solitonic solution could be presented as a modulation of harmonic
expression \eqref{2} of the corresponding mode
\cite{taniuti,oikawa,ramaz4}:
\begin{eqnarray}
\left(u_n^j, w_n^j\right)=\left(R_j,~~1\right)e^{i(p_jn-\Omega_j
t)}\varphi_j\left(\xi,\tau\right)+c.c. , \label{4}
\end{eqnarray}
\begin{figure}[t]
\includegraphics[scale=.60]{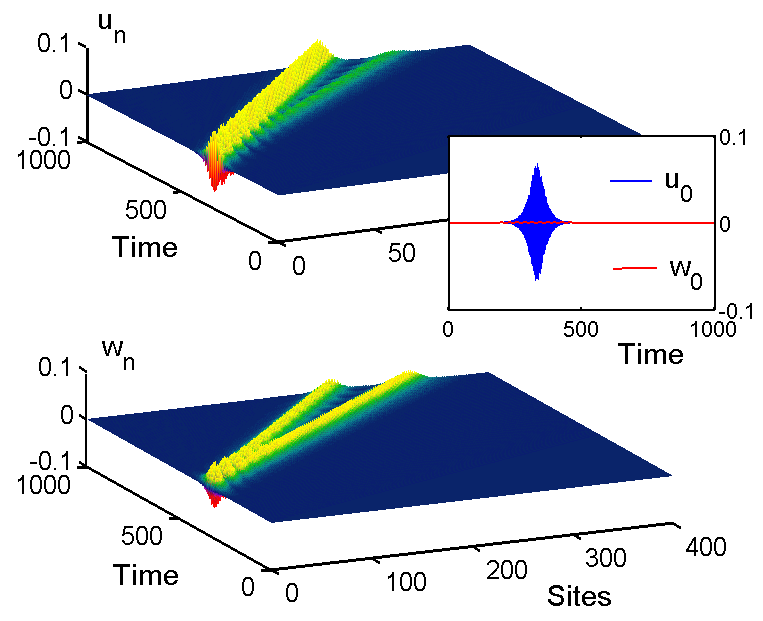}
\caption{Numerical simulations on the FPU ladder \eqref{1} with the
boundary driving according to \eqref{9}: the upper chain (upper
graph) is driven with a frequency $\Omega=1.9$ and the lower chain
(lower graph) is kept pinned. Inset shows how the ultimate left
$n=0$ balls of the ladder are driven in time.} \label{fig_3}
\end{figure}
where $\varphi_j(\xi,\tau)$ is a function of slow variables
$\xi=\epsilon(n-v_jt)$ and $\tau=\epsilon t^2$, where
$v_j=\partial\Omega_j(p)/\partial p\Bigr|_{p=p_j}$ is a group
velocity of the respective mode and $\epsilon$ is a small expansion
parameter. At the same time $\varphi_j(\xi,\tau)$ obeys the
nonlinear Schr\"{o}dinger (NLS) equation (please see for details SM
file):
\begin{eqnarray}
2i\frac{\partial\varphi_i}{\partial \tau}+\Omega_i ^{\prime
\prime}\frac{\partial^2\varphi_i}{\partial\xi^2}+\Delta_i|\varphi_i|^2\varphi_i=0
\label{5}
\end{eqnarray}
with the following parameters:
\begin{eqnarray}
\Delta_j=\frac{12(1-\cos p_j)^2(k_3+R_j^4)}{\Omega_j(1+R_j^2)},\quad
\Omega_j^{\prime\prime}={\frac{\partial^2\Omega_j}{\partial
p^2}}\Bigr|_{p=p_j}, \label{6}
\end{eqnarray}
and finally one arrives to the soliton solution of the respective
mode as follows:
\begin{eqnarray}
\left(u_n^j, w_n^j\right)=\left(R_j,~~1\right)\frac{A_j\cos(\Omega
t-p_jn)}{\cosh\left[(n-v_jt)/\Lambda_j\right]} \label{7}
\end{eqnarray}
where $A_j$ is a soliton amplitude, while soliton width $\Lambda_j$
and modified dispersion relation are given by:
\begin{eqnarray}
\Lambda_j=\frac{1}{A_j}\sqrt{\frac{2\Omega_j^{\prime\prime}}{\Delta_j}},
\qquad \Omega=\Omega_j+\frac{1}{4}\Delta_j A_j^2. \label{8}
\end{eqnarray}
Let us note that in the nonlinear case the latter relation has to be
applied for a computation of soliton carrier wavenumber $p_j$.
\begin{figure}[t]
\includegraphics[scale=.60]{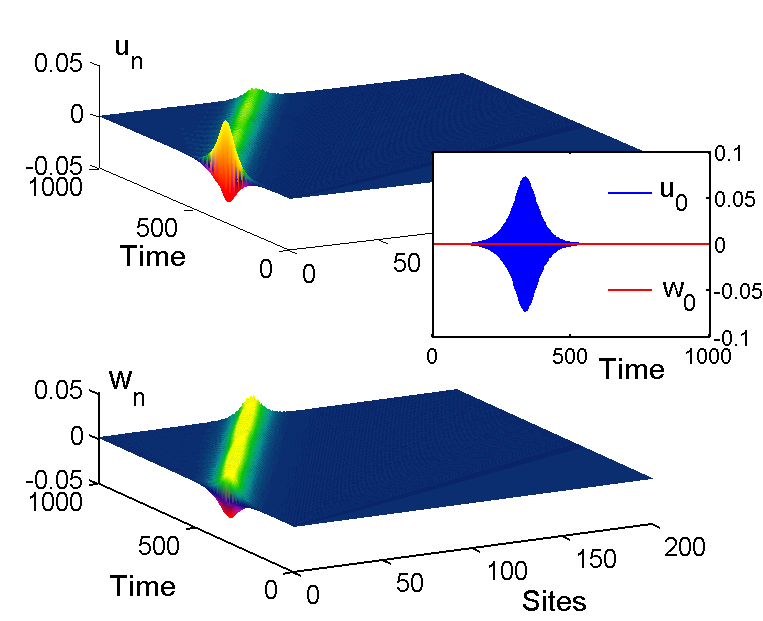}
\caption{Same as in Fig. 3, but now the upper chain is driven by a
frequency $\Omega=2.1$ and the lower chain is again pinned.}
\label{fig_4}
\end{figure}

In weakly nonlinear limit (small soliton amplitudes $A_j\ll 1$) and
large relative group velocities $\left|v_1-v_2\right|/v_{1,2}\gtrsim
1$ one can combine the solutions \eqref{7} acquiring additional
phase shift \cite{oikawa} which could be safely neglected in the
mentioned limits. By this one is able to construct the solution,
which describes the initial excitation of the boundary of the solely
upper chain. In particular, if one takes $A_1=A_2$ and finds such an
excitation frequency that $v_1/\Lambda_1=v_2/\Lambda_2$, the
combination $\left(u_n^1, w_n^1\right)-\left(u_n^2, w_n^2\right)$ at
the origin $n=0$ gives
\begin{eqnarray}
\left(u_0^1, w_0^1\right)-\left(u_0^2,
w_0^2\right)=\left(R_1-R_2,~~0\right)\frac{A_1\cos(\Omega
t)}{\cosh\left[v_1t/\Lambda_1\right]}, \label{9}
\end{eqnarray}
thus driving both chains in time according to the above expression
one can excite two soliton solution belonging to different branches.
That is displayed in Fig. 3, driving in numerical simulations the
left end of the upper chain $u_0$ with a frequency $\Omega=1.90$ and
amplitude $A=\left(R_1-R_2\right)A_1$ with $A_1=0.025$ and
calculating $R_j$ from Eq. \eqref{222}. At the same time the lower
chain is kept pinned at the left boundary ($w_0=0$) according again
to the expression \eqref{9}. As seen, the numerical test is just in
tact with the expectation, as  far as according to \eqref{7} we
observe different amplitudes for the solitons in the upper chain and
just the same $A_1=0.025$ in the lower one.

Next we examine one soliton generation driving again only upper
chain with a frequency lying in the limits
$\Omega_2(\pi)<\Omega<\Omega_1(\pi)$, particularly we apply
$\Omega=2.10$ in numerical simulations (see middle horizontal line
in Fig. 2). In this case antisymmetric mode ($j=1$) solution could
be again presented in solitonic form \eqref{7}, while the symmetric
mode ($j=2$) has no longer a solitonic profile, instead it is
described by evanescent wave since the corresponding wavenumber
$p_2$ is imaginary number (solution of dispersion relation
$\Omega=\Omega_2(p)$ has no real roots):
\begin{eqnarray}
\left(u_n^2,
w_n^2\right)=\left(R_2,~~1\right)B(t)e^{-\left|p_2\right|n}\cos(\Omega_2
t) \label{10}
\end{eqnarray}
where $B(t)$ can slowly vary in time. This means that we observe
only one soliton entering the chain. As we try to nullify
oscillations in the lower chain $B(t)$ should take a form of
$B(t)=A_1\sech\left(\emph{v}_1t/\Lambda_1\right)$ and then the
combination $\left(u_n^1, w_n^1\right)-\left(u_n^2, w_n^2\right)$ at
the origin $n=0$ gives the same form of the driving as in the
previous case \eqref{9} of the two soliton generation. The results
are displayed in Fig. 4, and as seen driving the upper chain with a
frequency $\Omega=2.10$ now one monitors the generation of a single
envelope soliton.

Finally we consider the case $\Omega=2.18$ (upper dashed line in
Fig. 2) lying in the band gap of both modes, for which only
evanescent wave solutions \eqref{10} is realized for the modes if
the driving amplitude is small. However, if the amplitude exceeds
some threshold value, a gap soliton can be created and propagate
along the ladder. For the estimation of this threshold value, we
assume that the upper chain is driven with the amplitude $A$ while
the lower one is kept pinned. Then, looking at the typical solution
of such a scenario \eqref{9} one can notice that the weight of the
antisymmetric mode $A_1$ is defined from the relation
$A=A_1\left(1-R_1/R_2\right)$ and the threshold value is calculated
from the expression of nonlinear frequency shift \eqref{8}:
\begin{eqnarray}
A^{th}=\left(1-R_1/R_2\right)\sqrt{4\left[\Omega-\Omega_1(\pi)\right]/\Delta_1}.
\label{11}
\end{eqnarray}

\begin{figure}[t]
\includegraphics[scale=.60]{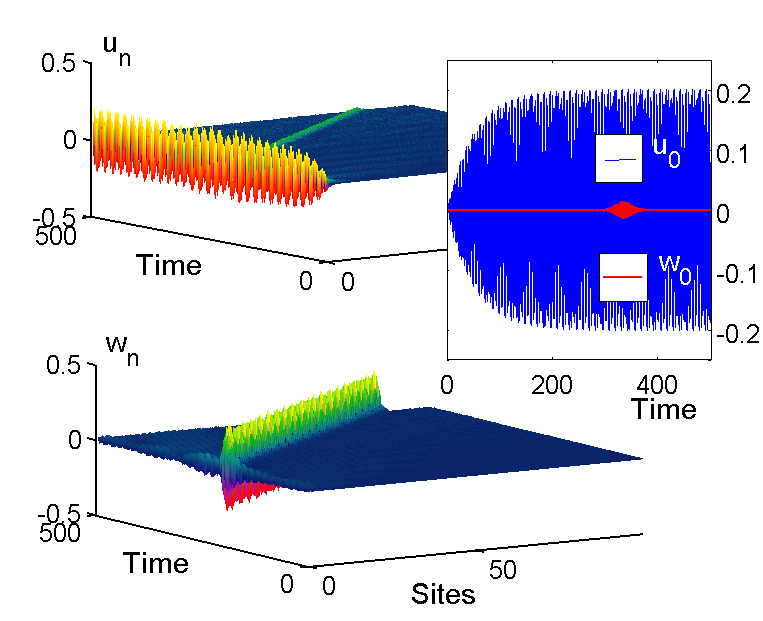}
\caption{The amplification scenario for FPU ladder. In the upper
chain (upper panel) we provide the continuous driving of the left
end with an amplitude just below the threshold \eqref{11} and the
band-gap frequency $\Omega=2.18$. While the left end of the lower
chain (lower panel) is perturbed by a small amplitude signal with
the same carrier frequency. Inset shows how the ultimate left $n=0$
balls of the ladder are driven in time.} \label{fig_5}
\end{figure}
Determining $A^{th}$ gives us an opportunity to realize the
amplification scenario. For this we create the continuous driving in
the upper chain with a band-gap frequency $\Omega=2.18$ and
amplitude just below the threshold, then even small counter-phase
pulse in the lower chain can help to overcome the threshold and
provide the necessary amplification effect for the weak pulse. For
the numerical experiment displayed in the Fig. 5 we use a continuous
driving with the amplitude $A=0.202$, while the pulse amplitude in
the lower chain can be of the order of $0.015$. As seen such a small
pulse is enough to create a gap soliton and realize amplification
scenario in the oscillator ladder. In order to provide a realistic
input-output ports we have lengthened lower chain adding linear part
at the left and nonlinear part at the right (see Fig. 1a) and as
could be seen the results are in agreement with developed analytical
scheme.

This amplification mechanism could be directly verified using
cantilever arrays \cite{sievers}, particularly, one can examine two
coupled in parallel cantilever arrays and use the scenario presented
in the Fig. 1. Note that although a cantilever array model include
onsite coupling terms in contrast to our model of the FPU lattices,
the consideration of the amplification mechanism will be the same,
since we consider upper band-gap localized modes (staggered
excitations) which are similar in both types of the anharmonic
chains. Using the soft mono-element lattice parameters from
\cite{sievers} and taking interchain coupling constant as $20\%$
from onsite coupling coefficient it follows that the upper band-gap,
appearing due to discreteness, starts at $\Omega_1(\pi)=116.8$KHz
and applying the driving in the band-gap with a frequency
$\Omega=117.8$KHz one can calculate the threshold amplitude for the
band-gap transmission according to \eqref{11} and it gives the value
$A_{th}=0.4\mu m$ that is much less than the typical distance
between cantilever and substrate $d=10\mu m$ and excitation can
propagate through the array. The width of the soliton appearing
after the amplification could be calculated from \eqref{8} and gives
the value $\Lambda_1\approx 10$ lattice sites. Thus it is very clear
that the developed amplification scheme is very robust with respect
to the model choice \cite{text}.

Concluding, it should be emphasized that we are using a single
operational frequency and thus the output signal could be readily
used for the further processing. Suggested mechanism could be
applied to study amplification in quantum systems, e.g. for trapped
cold atoms in optical lattice ladders. In this case engineering edge
defect site one can reach the threshold effect by tuning the atomic
onsite interaction strength. Then above some threshold value quantum
solitons will be created via resonance process between the edge
defect and the quantum bound state modes. But this issue needs a
further detailed investigation.

\begin{acknowledgments}
The work is supported in part by the grant from Georgian Shota
Rustaveli National Science Foundation (SRNSF) (Grant
\#FR/25/6-100/14) and joint grant from Science and Technology Center
in Ukraine (STCU) and SRNSF (Grant \#6084).
\end{acknowledgments}

\newpage

\section{Supplemental Material}

We start from the analysis of two weakly coupled FPU chains
displayed as Eq. (1) in the main text of the manuscript:
\begin{eqnarray}
m\ddot u_n &=&k_1^\prime(u_{n+1}+u_{n-1}-2u_n)+k_3^\prime(u_{n+1}-u_n)^3 \nonumber \\
&&
+k_3^\prime(u_{n-1}-u_n)^3+k(w_n-u_n) \nonumber \\
m\ddot w_n &=&k_1(w_{n+1}+w_{n-1}-2w_n)+k_3(w_{n+1}-w_n)^3 \nonumber
\\ && +k_3(w_{n-1}-w_n)^3+k(u_n-w_n) \label{101}
\end{eqnarray}

According to a well established procedure
\cite{taniuti,oikawa,ramaz4} of multi-scaling approach we are
seeking a weakly nonlinear solution of \eqref{101} in a form of
following perturbative expansion:
\begin{eqnarray}
\textbf{U}=\sum_{\alpha=1}^\infty\epsilon^\alpha
\sum_{m=-\infty}^{+\infty}{\textbf{U}_m^{(\alpha)}(\tau,\xi)e^{im(pn-\Omega
t)}} \label{exp}
\end{eqnarray}
where we define column vector
$\textbf{U}^{(\alpha)}=(u_n^{(\alpha)}, w_n^{(\alpha)})$, while
$\xi$ and $\tau$ are slow variables introduced through:
$\xi=\epsilon(n-vt)$ and $\tau=\epsilon t^2$; $v$ is a soliton group
velocity defined below and $\epsilon$ is a small expansion
parameter.

We go on with equating powers of $\epsilon$ substituting expansion
(\ref{exp}) in set of equations (\ref{1}). In the linear
approximation we have the column vector $\textbf{U}_1^{(1)}\equiv
\left( u_n^{(1)},w_n^{(1)}\right)=\varphi(\xi,\tau)\mathbf{R}$ and
$\textbf{U}_m^{(1)}=0$ for $|m|\neq1$; not restricting generality we
can take a space-time independent column vector as
$\mathbf{R}=\left(R,1\right)$, where $R$ is a complex number and
$\varphi(\xi,\tau)$ is a scalar function of slow variables to be
determined in the next approximations. Then by considering
$\alpha=1$ (linear approximation) and the harmonic $m=1$ we arrive
to the equation:
\begin{eqnarray}
\hat{\mathbf{W}}*\mathbf{R}=0 \label{WW}
\end{eqnarray}
where
\begin{eqnarray}\hat{\mathbf{W}}=\left(
\begin{array}{cc}
\Omega^{2}+2\left(\cos p-1\right)-k ~~~~~~~~~~~~~~ k~~~~~~~~~~~~~ \\
~~~~~~~~~~~k ~~~~~~~~~~~~~~ \Omega^{2}+2k_1\left(\cos p-1\right)-k
\end{array}
\right) \nonumber
\end{eqnarray}
the solvability of which demands Det($\hat{\mathbf{W}})=0$, which
gives us two branches of dispersion relations:
\begin{eqnarray}
 \Omega^2_{1,2}&=&(1-\cos p)(1+k_1)+k\pm \nonumber \\
&\pm&\sqrt{(1-\cos p)^2(1-k_1)^2+k^2} \label{disp}
\end{eqnarray}
and two corresponding column vectors $\mathbf{R}_j=(R_j,1)$ with
$R_j$ expressed with the linear parameters of the problem
$R_j=k/[k-\Omega^2_j+2(1-\cos p)]$, where $j=1,2$. Next we introduce
a row vector $\mathbf{L}=(L,1)$ through the equation
$\mathbf{L}*\hat{\mathbf{W}}=0$, that gives us two row vectors
$\mathbf{L}_j$. In our case the respective components of row
$\mathbf{L}_j$ and column $\mathbf{R}_j$ are identical $L_j=R_j$.
Thus in linear limit we have following matrix relations:
\begin{eqnarray}
\hat{\mathbf{W}}\left(\Omega_j\right)*\mathbf{R}_j=0, \qquad
\mathbf{L}_j*\hat{\mathbf{W}}\left(\Omega_j\right)=0. \label{WW1}
\end{eqnarray}
In the following for presentation clarity we omit the indexes $j$
and restore them at the end of the calculations.

We go on with a second approximation ($\alpha=2$) substituting again
\eqref{exp} into \eqref{101} and considering first harmonic $m=1$,
which leads us to the following equation:
\begin{eqnarray}
\hat{\mathbf{W}}\textbf{U}^{(2)}+2i(\hat{\mathbf{B}}-\Omega v
\hat{\mathbf{I}})\frac{\partial \varphi}{\partial \xi}\mathbf{R}=0,
\label{2222}
\end{eqnarray}
where
\begin{eqnarray} \hat{\mathbf{B}}=\left(
\begin{array}{cc}
\sin p & 0 \\ 0 & k_1\sin p
\end{array}
\right)
\end{eqnarray}

Then multiplying \eqref{2222} by $\mathbf{L}$ one has
\begin{eqnarray}
\mathbf{L}(\hat{\mathbf{B}}-\Omega v \hat{\mathbf{I}})\frac{\partial
\varphi}{\partial \xi}\mathbf{R}=0. \label{A3}
\end{eqnarray}
In order to identify constant $v$ in the equation above, let us take
the derivative of \eqref{WW} over $p$ and multiply then on the row
vector $\mathbf{L}$. One gets:
\begin{eqnarray}
\mathbf{L}\frac{\partial \hat{\mathbf{W}}}{\partial
p}\mathbf{R}=2\mathbf{L}\left(\frac{d\Omega}{dp}\Omega\hat{\mathbf{I}}-
\hat{\mathbf{B}}\right)\mathbf{R}=0. \label{A4}
\end{eqnarray}
Comparing now \eqref{A3} and \eqref{A4} we immediately get the
equality $v=\partial \Omega/\partial p$, thus the definition for
group velocity, while from \eqref{2222} one can solve
$\textbf{U}^{(2)}$ as follows:
\begin{eqnarray}
\textbf{U}^{(2)}=-2i\hat{\mathbf{W}}^{-1}(\hat{\mathbf{B}}-\Omega v
\hat{\mathbf{I}})\frac{\partial \varphi}{\partial \xi}\mathbf{R}.
\label{A5}
\end{eqnarray}

In the third approximation, equating powers of $\epsilon$ for
$\alpha=3$ and first harmonic $m=1$ we have:
\begin{eqnarray}
\hat{\mathbf{W}}\textbf{U}^{(3)}+2i(\hat{\mathbf{B}}-\Omega v
\hat{\mathbf{I}})\frac{\partial \textbf{U}^{(2)}}{\partial
\xi}+2i\Omega \frac{\partial\varphi}{\partial \tau}{\mathbf{R}}- \\
\nonumber
-\hat{\mathbf{C}}\frac{\partial^2\textbf{U}^{(1)}}{\partial\xi^2}+12(1-\cos
p)^2{\mathbf{N}}|\varphi|^2\varphi=0 \label{A6}
\end{eqnarray}
where
\begin{eqnarray}
\hat{\mathbf{C}} = \left(
\begin{array}{cc}
v^2-\cos p ~~~~~0 ~~~~~~~\\ ~~~0~~~~ v^2-k_1\cos p
\end{array} \right) ~~~
{\mathbf{N}} = \left(
\begin{array}{cc}
 R^3 \\ k_3
\end{array}
\right)
\end{eqnarray}

Now noting that
\begin{eqnarray}
2\left(\hat{\mathbf{B}}-\Omega v
\hat{\mathbf{I}}\right)\equiv-\frac{\partial
\hat{\mathbf{W}}}{\partial p}; \quad
\hat{\mathbf{C}}\equiv\frac{1}{2}\frac{\partial^2
\hat{\mathbf{W}}}{\partial p^2}-\Omega\frac{\partial^2
\Omega}{\partial p^2}\hat{\mathbf{I}} \label{A91}
\end{eqnarray}
We can further simplify \eqref{A6} multiplying it on $\mathbf{L}$
and taking into account \eqref{A5} and \eqref{A91}:
\begin{eqnarray}
\mathbf{L}\left(\Omega\frac{\partial^2 \Omega}{\partial
p^2}\hat{\mathbf{I}}+\frac{\partial \hat{\mathbf{W}}}{\partial
p}\hat{\mathbf{W}}^{-1}\frac{\partial \hat{\mathbf{W}}}{\partial
p}-\frac{1}{2}\frac{\partial^2 \hat{\mathbf{W}}}{\partial
p^2}\right)\textbf{R}\frac{\partial^2\varphi}{\partial \xi^2}
\nonumber \\
+2i\Omega \frac{\partial\varphi}{\partial \tau}\mathbf{L}\mathbf{R}
+12(1-\cos p)^2\mathbf{L}{\mathbf{N}}|\varphi|^2\varphi=0
\label{A61}
\end{eqnarray}

We can get a final form for \eqref{A61} taking first and second
derivatives of Eq. \eqref{WW} over $p$:
\begin{eqnarray}
\frac{\partial \hat{\mathbf{W}}}{\partial
p}\mathbf{R}+\hat{\mathbf{W}}\frac{\partial \textbf{R}}{\partial
p}=0; \label{A8} \\ \frac{\partial^2 \hat{\mathbf{W}}}{\partial
p^2}\mathbf{R}+2\frac{\partial \hat{\mathbf{W}}}{\partial
p}\frac{\partial \textbf{R}}{\partial
p}+\hat{\mathbf{W}}\frac{\partial^2 \textbf{R}}{\partial p^2}=0
\nonumber
\end{eqnarray}
Solving now $\partial \textbf{R}/\partial p$ from the first equation
and substituting it in the second one and then multiplying it on
$\textbf{L}$ one gets the following relation:
\begin{eqnarray}
\mathbf{L}\frac{\partial \hat{\mathbf{W}}}{\partial
p}\hat{\mathbf{W}}^{-1}\frac{\partial \hat{\mathbf{W}}}{\partial
p}\textbf{R}-\frac{1}{2}\mathbf{L}\frac{\partial^2
\hat{\mathbf{W}}}{\partial p^2}\mathbf{R}=0, \label{A9}
\end{eqnarray}
and now substituting this into the \eqref{A61} and restoring indexes
$j$-s one finally arrives to the Nonlinear Schr\"odinger (NLS)
Equation for two nonlinear modes $j=1,2$:
\begin{eqnarray}
2i\frac{\partial\varphi_j}{\partial \tau}+\Omega_j^{\prime
\prime}\frac{\partial^2\varphi_j}{\partial\xi^2}-\Delta_j|\varphi_j|^2\varphi_j=0
\label{nls}
\end{eqnarray}
where
\begin{eqnarray}
\Delta_j=\frac{12(1-\cos p_j)^2(k_3+R_j^4)}{\Omega_j(1+R_j^2)},\quad
\Omega_j^{\prime\prime}={\frac{\partial^2\Omega_j}{\partial
p^2}}\biggr|_{p=p_j} \label{nls0}
\end{eqnarray}
and wavenumbers $p_j$ are the solutions of respective dispersion
relations $\Omega=\Omega_j+\Delta_j A_j^2/4$.

We use the same approach considering cantilever arrays. Beginning
with modified equations of motion:
\begin{eqnarray}
&&m\ddot u_n +k_{20}^\prime u_n+k_{40}^\prime u_n^3 -k_1^\prime(u_{n+1}+u_{n-1}-2u_n)- \nonumber \\
&&-k_3^\prime(u_{n+1}-u_n)^3-k_3^\prime(u_{n-1}-u_n)^3+k(u_n-w_n)=0 \nonumber \\
&&m\ddot w_n+k_{20}u_n+k_{40}u_n^3-k_1(w_{n+1}+w_{n-1}-2w_n)-
\nonumber\\
&&~k_3(w_{n+1}-w_n)^3-k_3(w_{n-1}-w_n)^3+k(w_n-u_n)=0 \nonumber
\end{eqnarray}
In order to estimate the effect we take the following approximate
values of the problem parameters: $m=10^{-12}kg$,
$k_1=k_1^\prime=k_{20}=k_{20}^\prime=0.1kg/s^2$,
$k_3=k_3^\prime=10^{10}kg/s^2m^2$,
$k_{40}=k_{40}^\prime=10^8kg/s^2m^2$, and take weak interchain
linear coupling coefficient as $k=0.02kg/s^2$. Then the frequencies
of two branches $\Omega_1$ and $\Omega_2$ are the solutions of
matrix dispersion relation
\begin{eqnarray}
Det\left[
\begin{array}{cc}
\Omega^{2}+2k_1^\prime\left(\cos p-1\right)-k-k_{20}^\prime ~~~~~~~~~~ k~~~~~~~~ \\
~~~~~~~~k ~~~~~~~~~ \Omega^{2}+2k_1\left(\cos p-1\right)-k-k_{20}
\end{array}
\right]=0 \nonumber
\end{eqnarray}
and from the ordinary procedure developed above we again get NLS
equations with following nonlinear coefficients for the
antisymmetric ($j=1$) and symmetric ($j=2$) modes:
\begin{eqnarray}
\Delta_j=\frac{12(1-\cos p_j)^2(k_3+k_3^\prime
R_j^4)+k_{40}+k_{40}^\prime R_j^4}{\Omega_jk_3^\prime(1+R_j^2)}
\nonumber
\end{eqnarray}
From this point one can obtain the value of the band-gap frequency
and the threshold amplitude as shown above. In our case
$\Omega_1(\pi)=116.8$KHz and $A_{th}=0.4\mu m$ with the driving
frequency $\Omega=117.8$KHz.


\begin{thebibliography}{99}
\bibitem{text} Another option is to use a single array realization by considering
two-dimensional vibrations of crystalline pillars \cite{qart} where
one can achieve the amplification of the longitudinal acoustic wave
via driving transversal vibrations or vice versa.
\end{thebibliography}
\end{document}